\documentclass {elsart}
\usepackage[english]{babel}
\usepackage {amssymb,graphicx,subfigure}
\usepackage{cite}
\usepackage{amsmath}

\def\anti#1{\mathpalette{\@anti}{#1}#1}
\def\@anti#1#2{\sbox0{$#1#2$}
  \makebox[0pt][l]
    {$#1\kern.30\ht0\overline{\kern-.35\ht0\phantom{#2}\kern-.1ex}$}}

\begin{document}
\begin{frontmatter}

\title{
 \bf Study of the doubly and singly Cabibbo suppressed
 decays
 $D^+ \to K^+ \pi^+ \pi^-$ and
 $D^+_s \to K^+ \pi^+ \pi^-$
 }
 The FOCUS Collaboration
\author[ucd]{J.~M.~Link}
\author[ucd]{P.~M.~Yager}
\author[cbpf]{J.~C.~Anjos}
\author[cbpf]{I.~Bediaga}
\author[cbpf]{C.~G\"obel}
\author[cbpf]{A.~A.~Machado}
\author[cbpf]{J.~Magnin}
\author[cbpf]{A.~Massafferri}
\author[cbpf]{J.~M.~de~Miranda}
\author[cbpf]{I.~M.~Pepe}
% ...

\address[ucd]{University of California, Davis, CA 95616}
\address[cbpf]{Centro Brasileiro de Pesquisas F\'isicas, Rio de Janeiro, RJ, Brasil}
\address[cinv]{CINVESTAV, 07000 M\'exico City, DF, Mexico}
\address[cu]{University of Colorado, Boulder, CO 80309}
\address[fnal]{Fermi National Accelerator Laboratory, Batavia, IL 60510}
\address[fras]{Laboratori Nazionali di Frascati dell'INFN, Frascati, Italy I-00044}
\address[guan]{University of Guanajuato, 37150 Leon, Guanajuato, Mexico}
\address[ui]{University of Illinois, Urbana-Champaign, IL 61801}
\address[iu]{Indiana University, Bloomington, IN 47405}
\address[korea]{Korea University, Seoul, Korea 136-701}
\address[korea2]{Kyungpook National University, Taegu, Korea 702-701}
\address[milan]{INFN and University of Milano, Milano, Italy}
\address[nc]{University of North Carolina, Asheville, NC 28804}
\address[pavia]{Dipartimento di Fisica Nucleare e Teorica and INFN, Pavia, Italy}
\address[pr]{University of Puerto Rico, Mayaguez, PR 00681}
\address[sc]{University of South Carolina, Columbia, SC 29208}
\address[ut]{University of Tennessee, Knoxville, TN 37996}
\address[vu]{Vanderbilt University, Nashville, TN 37235}
\address[wisc]{University of Wisconsin, Madison, WI 53706}

\endnote{\small
  See \textrm{http://www-focus.fnal.gov/authors.html} for additional author information.
}

\begin{abstract}

Using data collected by the high energy photoproduction experiment FOCUS at
Fermilab we study the doubly and singly Cabibbo suppressed decays $D^+$ and $D^+_s
\to K^+ \pi^+ \pi^-$. Our measurements of $\Gamma(D^+\to K^+ \pi^+
\pi^-)/\Gamma(D^+\to K^- \pi^+ \pi^+)=0.0065 \pm 0.0008 \pm
0.0004$ and $\Gamma(D^+_s\to K^+ \pi^+\pi^-)$/ $\Gamma(D^+_s\to
K^+ K^- \pi^+)=0.127 \pm 0.007\pm 0.014$ are based on samples of $189\pm24$
$D^+$ and $567 \pm 31$ $D^+_s$ reconstructed events, respectively. We also present
Dalitz plot analyses of the two decay channels; the amplitude analysis of the
$D^+_s \to K^+ \pi^+ \pi^-$ mode is performed for the first time.

\end{abstract}

\end{frontmatter}


\section{Introduction}

A thorough understanding of $D$ hadronic decays requires both 
branching ratio measurements
of all the hadronic modes 
and Dalitz plot 
analyses to investigate the decay dynamics and probe the role of 
final state interactions. 
In this letter we present branching ratio measurements and Dalitz plot analyses of the 
doubly and singly Cabibbo suppressed decays $D^+ \to K^+ \pi^+ \pi^-$ and
$D^+_s \to K^+ \pi^+ \pi^-$.

The na\"ive expectation for the ratio of Doubly Cabibbo Suppressed (DCS)
to Cabibbo Favored (CF) branching fractions is $\tan^4\theta_\text{C} \sim 0.25 $\%.
However the CF $D^+ \to K^- \pi^+ \pi^+$ rate could be suppressed by
destructive interference between spectator amplitudes with indistinguishable
quarks in the final state; this is the argument generally proposed to explain
the lifetime difference between $D^+$ and $D^0$. For the hadronic DCS $D^+$
decay all the final state quarks are different and no destructive interference
is present. In this simple picture we would expect, neglecting effects of final
state interactions, $\tau(D^+)/\tau(D^0) =
\Gamma(D^0_{\text{CF}})/\Gamma(D^+_{\text{CF}})
=(1/\tan^4\theta_\text{C}) \times \Gamma(D^+_{\text{DCS}})/\Gamma(D^+_{\text{CF}})$. The comparison of
the precise FOCUS lifetime ratio \cite{focus_life} $\tau(D^+)/\tau(D^0)=2.538
\pm 0.023$ with the branching ratio $\Gamma(D^+_{\text{DCS}})/\Gamma(D^+_{\text{CF}})$
measurement reported here will test this interpretation.

The final $D^+$ and $D^+_s$ samples are selected with cuts to reduce
reflections from the more copious Cabibbo favored modes 
and to
optimize the signal to noise ratios, which are crucial for reliable decay
amplitude analyses. 
Dalitz plot analyses have indeed emerged as a unique tool to
fully exploit the available charm statistics allowing, besides the simple
branching ratio evaluation, to investigate the underlying decay dynamics.
Our understanding of the charm decay dynamics has already considerably 
improved in the last years, but is still limited to a few decay channels.
The Dalitz plot analysis of the Singly Cabibbo Suppressed decay (SCS)
$D^+_s\to K^+ \pi^+\pi^-$ is performed for the first time.
%

\section{Signal selection}

The data for this analysis were collected during the 1996--1997 run of the
photoproduction experiment FOCUS at Fermilab. The detector, designed and used
to study the interaction of high-energy photons on a segmented BeO target, is a
large aperture, fixed-target magnetic spectrometer with excellent \v{C}erenkov
particle identification and vertexing capabilities. Most of the FOCUS
experiment and analysis techniques have been described previously
\cite{E687_spectr,Focus_cherenkov,life_diff}.

The suppressed nature of the channels under study requires severe
cuts to eliminate reflections, both in the signal and sideband regions,
from more copious and favored decays, when one or two charged particles are
\v{C}erenkov misidentified and/or one neutral particle is missing. Tight cut
choices are also required to improve the signal-to-noise ratio 
to perform a reliable Dalitz plot analysis. 

The $D^+$ and $D^+_s$ candidates are obtained
using a candidate driven vertex algorithm. A decay vertex is formed from three
reconstructed charged tracks. The momentum of the $D$ candidate is used to
intersect other reconstructed tracks to form a production vertex. The
confidence levels (C.L.) of each vertex is required to exceed 1\%. From the
vertexing algorithm, the variable $\ell$, which is the separation of the
primary and secondary vertices, and its associated error $\sigma_\ell$ are
calculated. We reduce backgrounds by requiring $\ell/\sigma_\ell>14$ and 10 for
the $D^+$ and $D_s^+$, respectively. The two vertices are also required to
satisfy isolation conditions. The primary vertex isolation cut requires that a
track assigned to the decay vertex has  a C.L. less than 1\% to be included
in the primary vertex. The secondary vertex isolation cut requires that all
remaining tracks not assigned to the primary and secondary vertex have a C.L.\
smaller than 0.1\% to form a vertex with the $D$ candidate daughters. The
decay vertex is required to be  $3\,\sigma$ outside of the target material to
reduce the background due to hadronic re-interactions in the target.

The \v{C}erenkov particle identification is based on likelihood ratios between
the various stable particle hypotheses \cite{Focus_cherenkov}. The product of
all firing probabilities for all cells within the three \v{C}erenkov cones
produces a $\chi^2$-like variable $W_i$=-2ln(likelihood) where $i$ ranges over
the electron, pion, kaon and proton hypothesis. For the $D^+$ selection, we
require $\Delta W_K= W_{\pi} -W_{K}$ 
$> 4$ and $\Delta
W_{\pi}= W_{K} -W_{\pi}$ 
$> 3.5$ for both pions in the final
state. For the $D^+_s$ candidates we require $\Delta W_K > 4$ for the
kaon, $\Delta W_{\pi}> 2$ for opposite-sign pion and $\Delta W_{\pi}> 1$ for
same-sign pion. These selections minimize the contamination from charm
background.

\begin{figure}[!t]
 \begin{center}
  \subfigure[]
  {
  \includegraphics[width=0.45\textwidth]{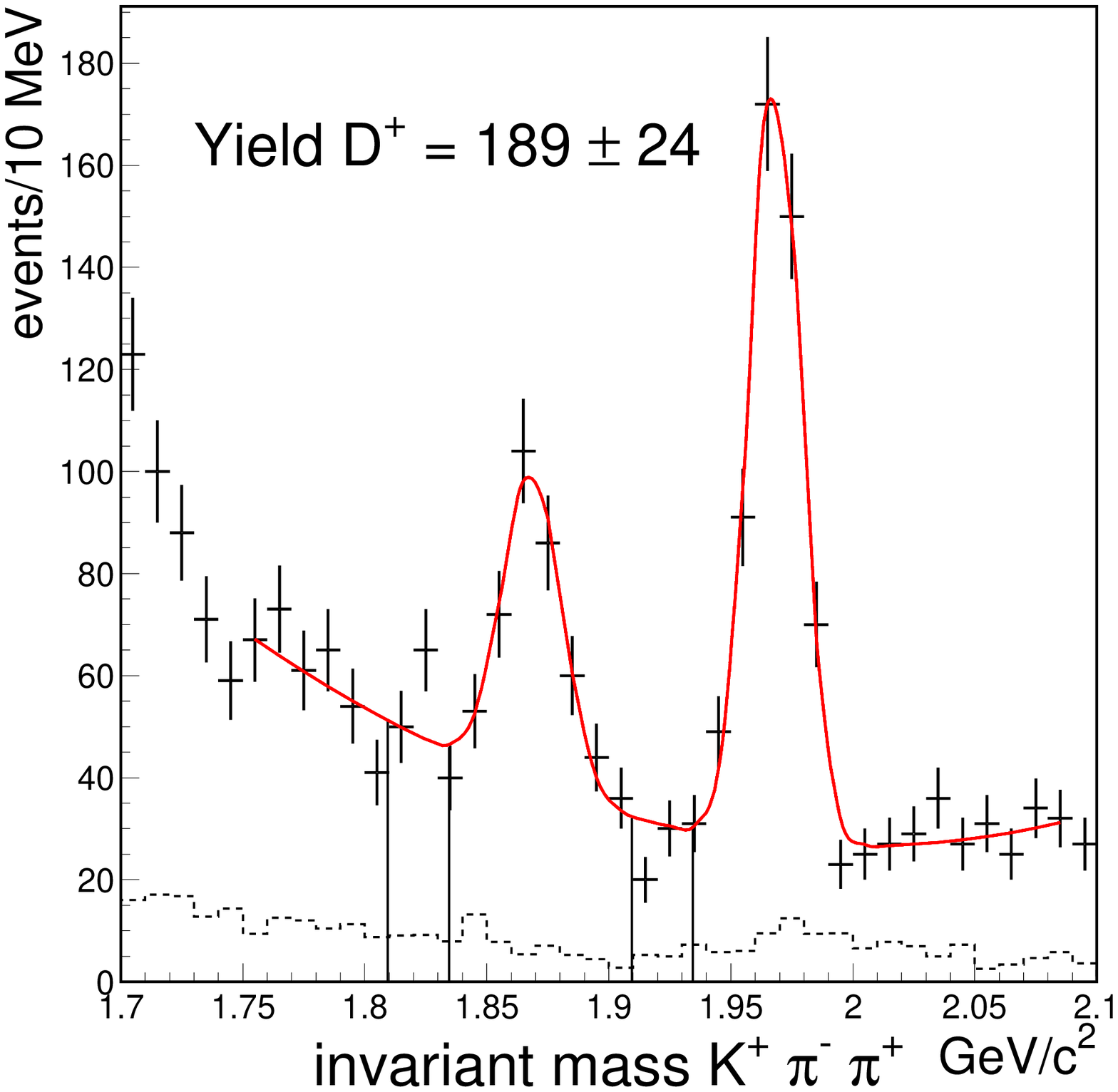}
  \label{segnale_dcs}
  }
   \subfigure[]
  {
   \includegraphics[width=0.45\textwidth]{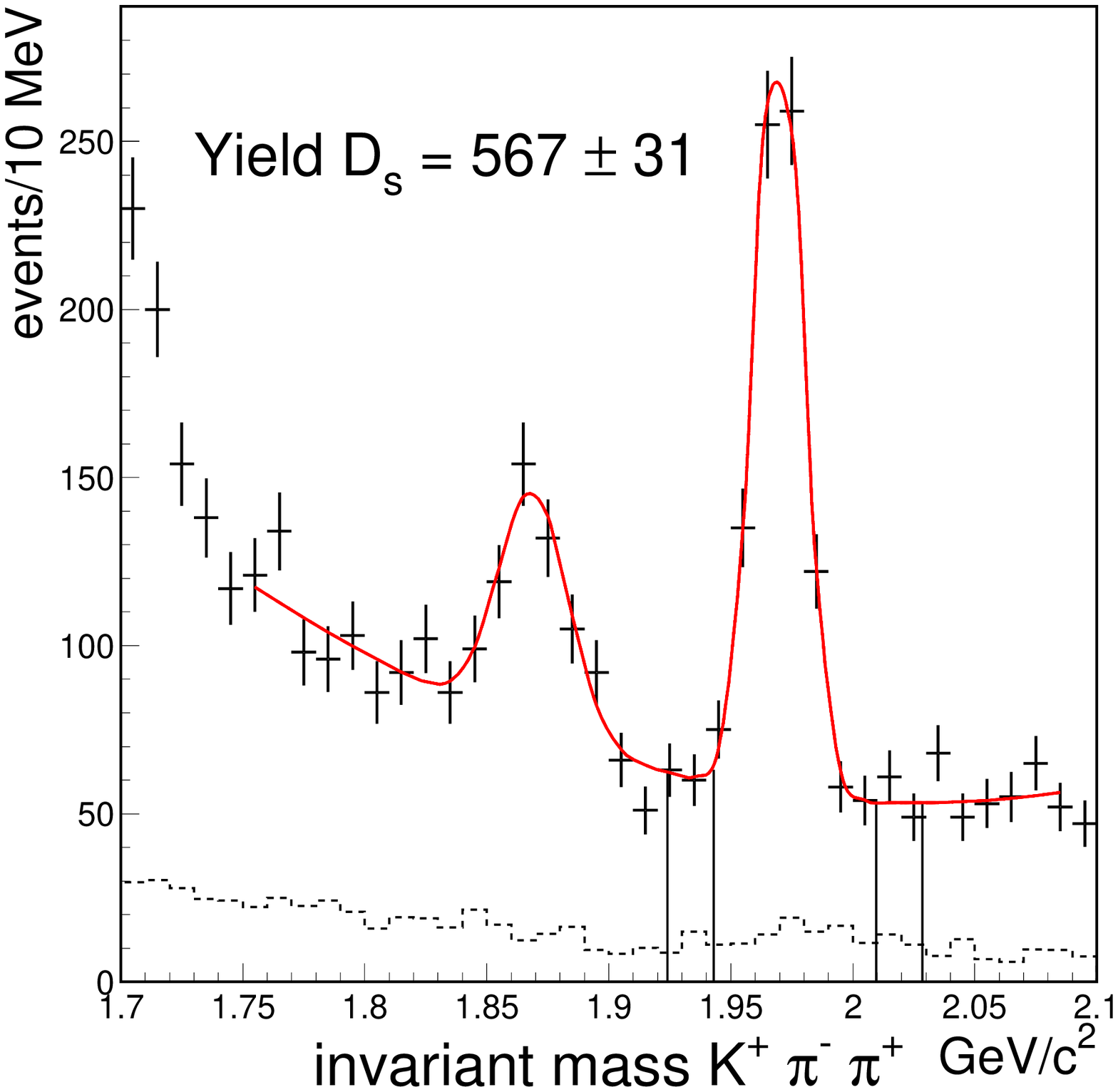}
  \label{segnale_scs}
  }
 \end{center}
 \caption{Invariant mass distributions $K^+ \pi^+ \pi^-$ with final
 selection criteria for (a) $D^+$ and (b) $D^+_s$, along with distributions of
  the residual reflected events expected from Monte Carlo (dotted curves). In the 
  picture the sideband regions (vertical lines) are also shown. }
 \label{segnali}
\end{figure}

A detailed Monte Carlo study is performed to evaluate the possible contaminations
and structures induced by reflections both in the signal and sideband regions.\footnote{The 
sidebands for $D^+$ cover the $-5\sigma$ to $-3\sigma$ and the
$3\sigma$ to $5\sigma$ regions from the $D^+$ peak. For the $D_s^+$ the left
sideband covers the $-5\sigma$ to $-3\sigma$ region and the right sideband
covers the $4\sigma$ to $6\sigma$ region, both from the $D_s^+$ peak.} The
considered sources of background are the $D^+$ and $D^+_s$ three-body hadronic
decays, where one or two particles are misidentified, and four-body hadronic
and semileptonic decays, where charged particles are misidentified and neutrals
are missing, namely: $D^+ \to K^- \pi^+ \pi^+$, $D^+_s \to K^- K^+ \pi^+$, $D^+
\to K^- K^+ \pi^+$, $D^+ \to \pi^+ \pi^- \pi^+$,  $D^+ \to K^- \pi^+ \mu^+
\nu$, $D^+_s \to K^- K^+ \mu^+ \nu$ and $D^+ \to K^- \pi^+ \pi^+ \pi^0$. For
each of them we evaluate the number and the distribution shape of events which
survive the selection cuts and penetrate into the $D^+$ and $D^+_s
\to K^+ \pi^+ \pi^-$ signal and sideband regions.
From this study, we find that the major contributions for $D^+$ come from
$D^+ \to K^- \pi^+ \pi^+$, $D^+_s \to K^- K^+ \pi^+$ and $D^+ \to K^-
\pi^+\pi^+\pi^0$, and for $D_s^+$ from $D^+ \to K^- \pi^+ \pi^+$, $D^+ \to
\pi^+\pi^-\pi^+$ and $D^+ \to K^- \pi^+\pi^+\pi^0$. In Fig. \ref{segnali} the
selected $K^+\pi^+\pi^-$ combinations for $D^+$ and $D^+_s$ are shown, along with
the distribution of the Monte Carlo reflected events. We choose to start
our fit from the 1.75 $\textrm{GeV}/c^2$ energy threshold since the region below is
dominated by the partial reconstruction of multi-body charm channels where
neutrals are missing. The signal yields consist of $189 \pm 24$ events for $D^+$
and $567 \pm 31$ events for $D^+_s$. The fits are performed with two Gaussian
functions for the signals and a second order polynomial for the background.
Centroid and width are $1.869 \pm 0.002 ~\textrm{GeV}/c^2$ and $0.012 \pm 0.002 ~\textrm{GeV}/c^2$
for $D^+$ (Fig. \ref{segnale_dcs}), and $1.970 \pm 0.001 ~\textrm{GeV}/c^2$ and $0.010 \pm
0.001 ~\textrm{GeV}/c^2$ for $D^+_s$ (Fig. \ref{segnale_scs}). The measured widths are in
good agreement with the Monte Carlo predictions. The signal to noise ratio are
$1.0 \pm 0.1$ and $2.4 \pm 0.4$ for $D^+$ and $D^+_s$, respectively. 
The reflected events are smoothly
distributed across the invariant mass spectrum and are thus properly accounted for
by the background polynomial fitting function.

Residual structures induced by reflections in the Dalitz plot signal and sideband
regions are a potential source of systematics of our amplitude analysis. The shape
of the background in the Dalitz signal region is parametrized through a fit to the
Dalitz plot of mass sidebands, which consists of a polynomial plus Breit-Wigner
functions to account for any feed-through from resonances. We verify that the
reflections in the sidebands are well described by this simple background fit
function and adequately represent the reflections in the signal regions. 
%

\section{Branching ratio results}

\begin{figure}[!t]
 \begin{center}
  {
   \includegraphics[width=0.45\textwidth]{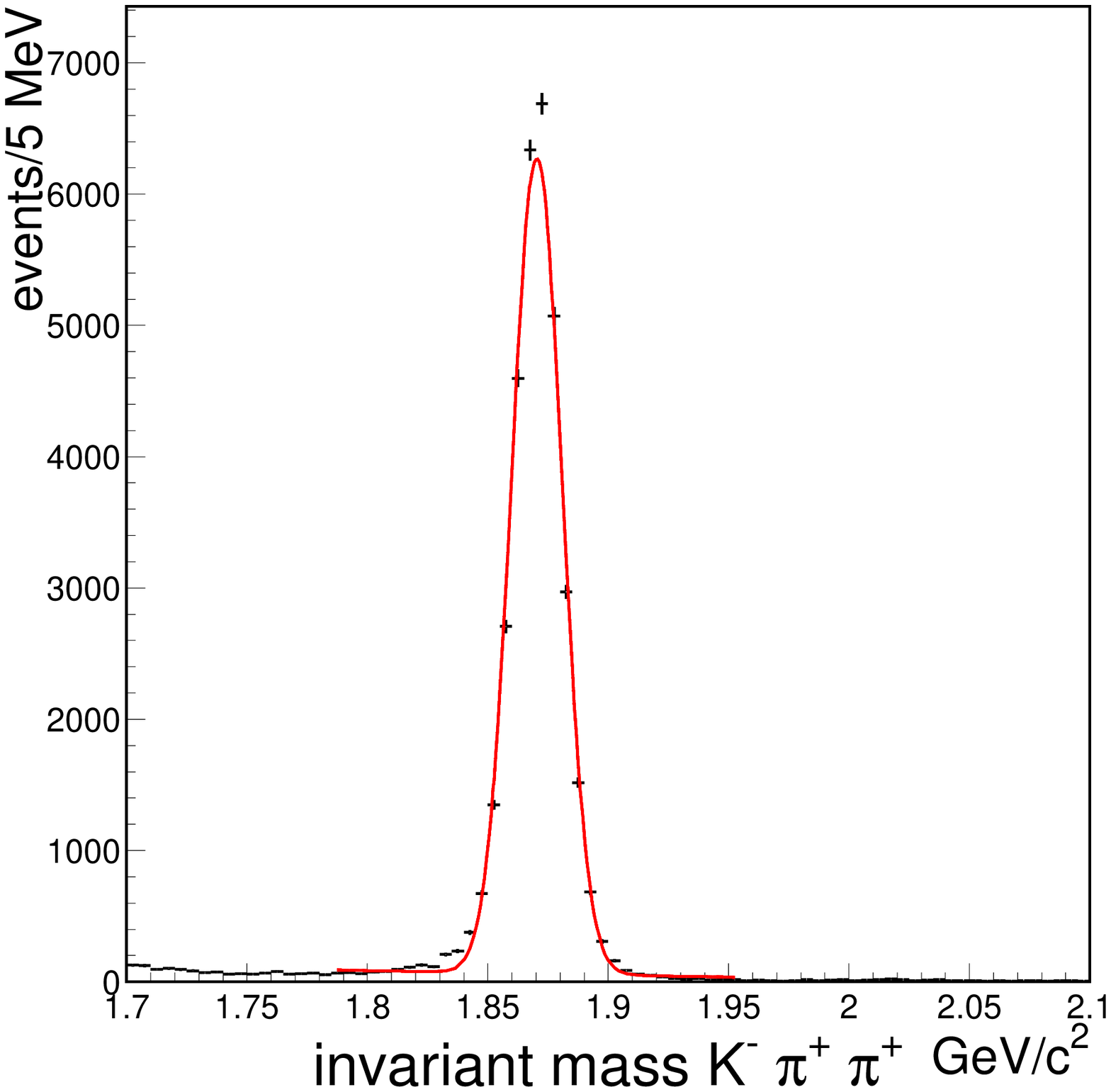}
  \label{segnale_cf}
  }
  {
   \includegraphics[width=0.45\textwidth]{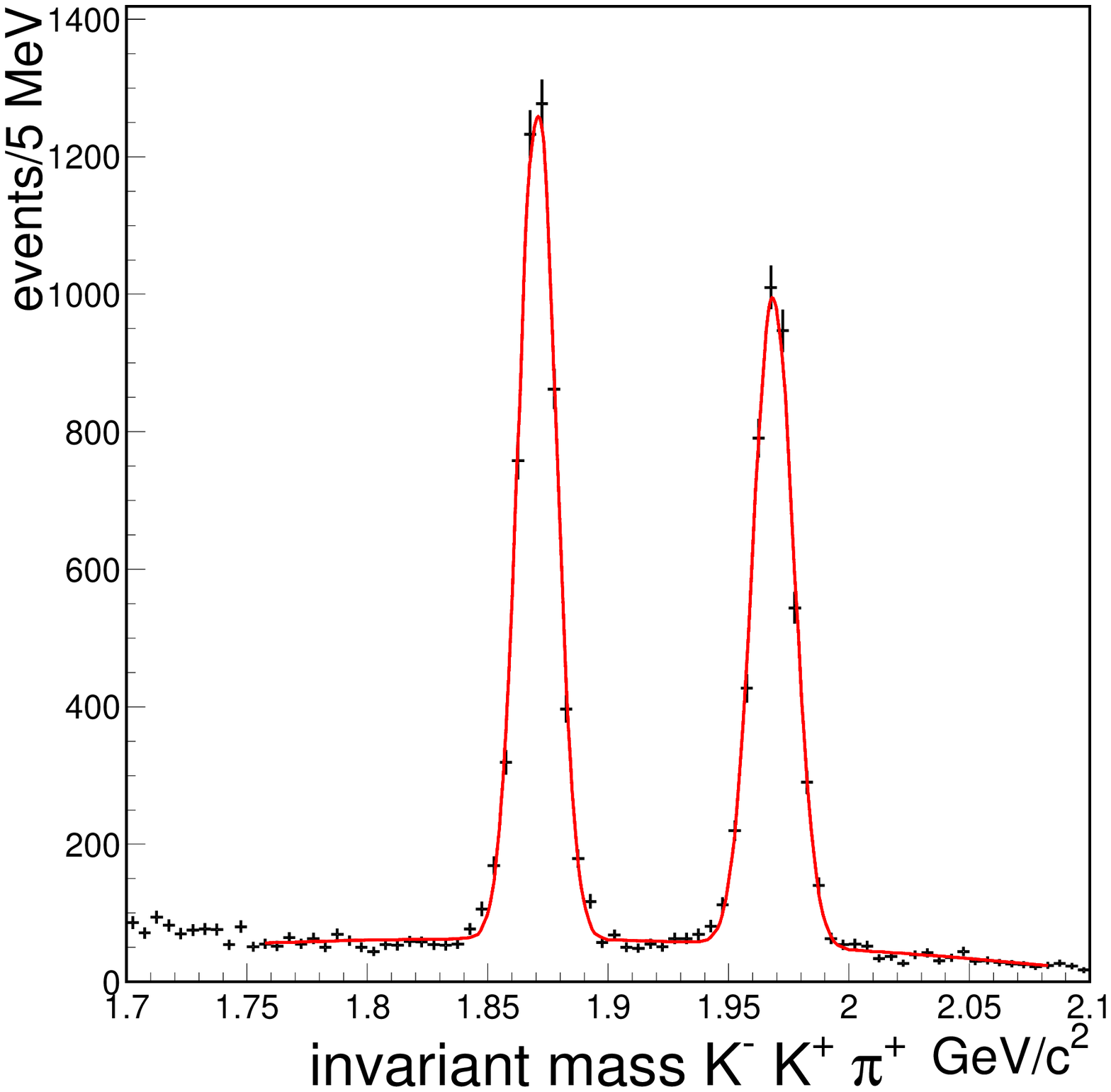}
  \label{segnale_kkp}
  }
  \caption{Invariant mass distributions of $D^+\to K^- \pi^+\pi^+$ and
  $D^+_s \to K^- K^+ \pi^+$ normalization channels.}
  \label{segnale_norm}
 \end{center}
\end{figure}

The DCS $D^+ \to K^+ \pi^+ \pi^-$ decay fraction is evaluated with respect to the CF
$D^+ \to K^- \pi^+ \pi^+$ mode, the SCS $D^+_s \to  K^+ \pi^+ \pi^-$ with respect to
the CF $D^+_s \to K^- K^+ \pi^+$ mode.
 In Fig. \ref{segnale_norm} the mass distributions for the normalization
channels are shown. The CF  $D^+ \to K^- \pi^+ \pi^+$ sample is selected with
the same set of cuts as for the DCS channel, in order to minimize systematic
effects; the signal yield is $32714 \pm 184$.
The CF $D^+_s\to K^+ K^- \pi^+$ is obtained requiring for $K^-$ a $\Delta W_K > 2$
similar to the $\Delta W_\pi$ for the opposite-sign pion in the
SCS selection; kinematical cuts are applied to remove the reflections in the
$K^-K^+\pi^+$ mode from the CF $D^+ \to K^- \pi^+ \pi^+$ and from $D^{*+} \to
D^0 \pi^+$, followed by the $D^0$ decay to $K^- \pi^+$. The $D^+_s \to K^- K^+
\pi^+$ signal yield is $4033 \pm 68$.
A fully coherent generation is used both for the channels under study and the 
normalization modes; the relative efficiencies are measured to be
\begin{equation}
\frac{\epsilon(D^+ \to K^+ \pi^+ \pi^-)}{\epsilon(D^+\to K^- \pi^+\pi^+)} = 0.888
\pm 0.006 
\end{equation}
and 
\begin{equation}
\frac{\epsilon(D^+_s \to K^+ \pi^+ \pi^-)}{\epsilon(D^+_s\to K^- K^+\pi^+)} = 1.106
\pm 0.009 \, .
\end{equation}

From the yields and efficiency ratios, we measure the following branching ratios,
\begin{equation}
\frac {\Gamma\left(D^+\to K^+ \pi^+ \pi^-\right)} {\Gamma\left(D^+\to K^- \pi^+
\pi^+\right)} = 0.0065 \pm 0.0008 \pm 0.0004
\end{equation}
and
\begin{equation}
\frac {\Gamma\left(D^+_s\to K^+ \pi^+ \pi^-\right)} {\Gamma\left(D^+_s\to K^+ K^-
\pi^+\right)} = 0.127 \pm 0.007 \pm 0.014 \, .
\end{equation}

The first error reported is statistical and the second is systematic.

To evaluate the systematic error we consider, as done in other
FOCUS analyses \cite{sistematica1,sistematica2}, three contributions, which are added in
quadrature to obtain the global systematic error: the \emph{split sample}, 
\emph{fit variant} and \emph{cut variant} components.
The \emph{split sample} component takes into account the systematics introduced by residual
difference between data and Monte Carlo, due to a possible mismatch in the reproduction of the $D$
momentum and the changing experimental conditions of the spectrometer during data collection.
This component has been determined by splitting data into four independent subsamples, according
to the $D$ momentum range (high and low momentum) and the configuration of the vertex detector.
The
\emph{S-factor method} from the Particle Data Group \cite{pdg} was used to try to separate true
systematic variations from statistical fluctuations. The branching ratio is evaluated for each of
the four statistically independent subsamples and a \emph{scaled variance} is calculated
$\widetilde{\sigma}$ (the errors are boosted when $\chi^2/(N-1)>1$). The \emph{split sample}
variance is defined as the difference between the reported statistical variance and the scaled
variance, if the scaled variance exceeds the statistical variance.
Another possible source of systematics uncertainty is the \emph{fit variant}. This component is
computed by varying the fitting conditions on the whole data set.
In our study fit variants include the background and signal shape, the bin size of the 
mass-distribution histogram and the Monte Carlo generation modeling.
More precisely different degrees of the polynomial functions are used for the 
background parametrization and two Gaussian peaks with 
the same mean but different widths are used for the signal, to take into account 
the different resolution in momentum of our spectrometer. 
The fully coherent generation for the decays under analysis is based on the 
results of the Dalitz plot study presented in this paper (see paragraph \ref{results}).
To access a possible systematics effect in the branching ratio evaluation 
coming from our amplitude analysis we vary coefficients and phases returned by the 
Dalitz plot fit within their errors.
The branching ratio values obtained by these variants are all a priori likely, therefore this
uncertainty can be estimated by the standard deviation of the measurements. 
To investigate the dependence of the branching ratio on the particular choice 
of cuts we calculate a 
\emph{cut variant} error, analogously to the \emph{fit variant}, by using the standard deviation 
of different branching ratio measurements obtained with several sets of cuts; 
this error is actually overestimated since the statistics of the cut samples are 
different.

For the $D^+$ the dominant systematic effect comes from the \emph{cut variant},
 while for the $D^+_s$
the main source of uncertainty comes from the \emph{split sample} component. Our
results are consistent with previous determinations from the E687 \cite{E687_DCS}
and E791 \cite{E791} experiments. Our measurements reduce the statistical errors
by about a factor of 2 and 5 for the $D^+$ and $D^+_s$, respectively.
%

\section{Dalitz plot analysis}

\begin{figure}[!t]
 \begin{center}
 \subfigure[]
  {
  \includegraphics[width=0.47\textwidth]{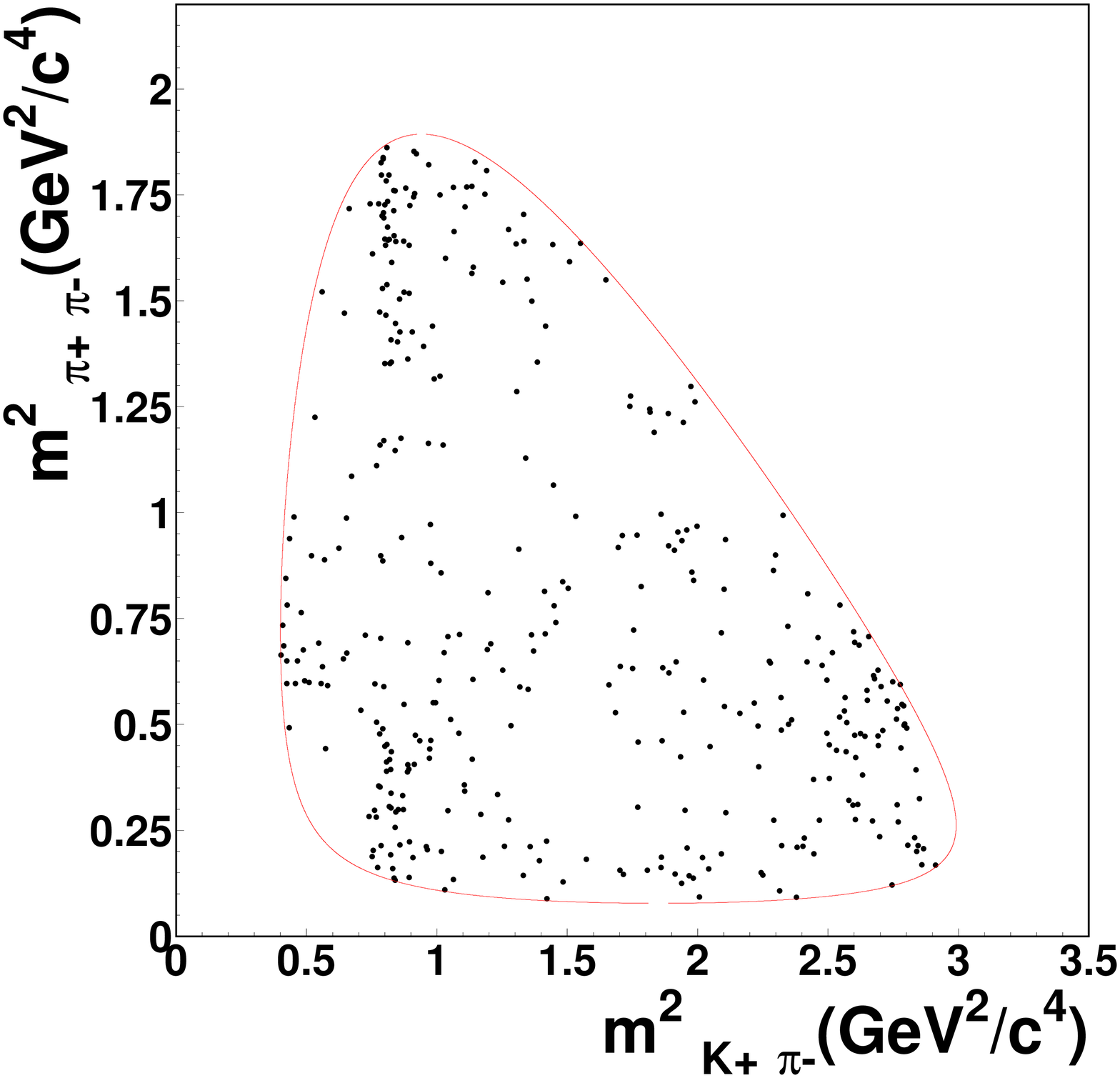}
  }
 \subfigure[]
  {
  \includegraphics[width=0.47\textwidth]{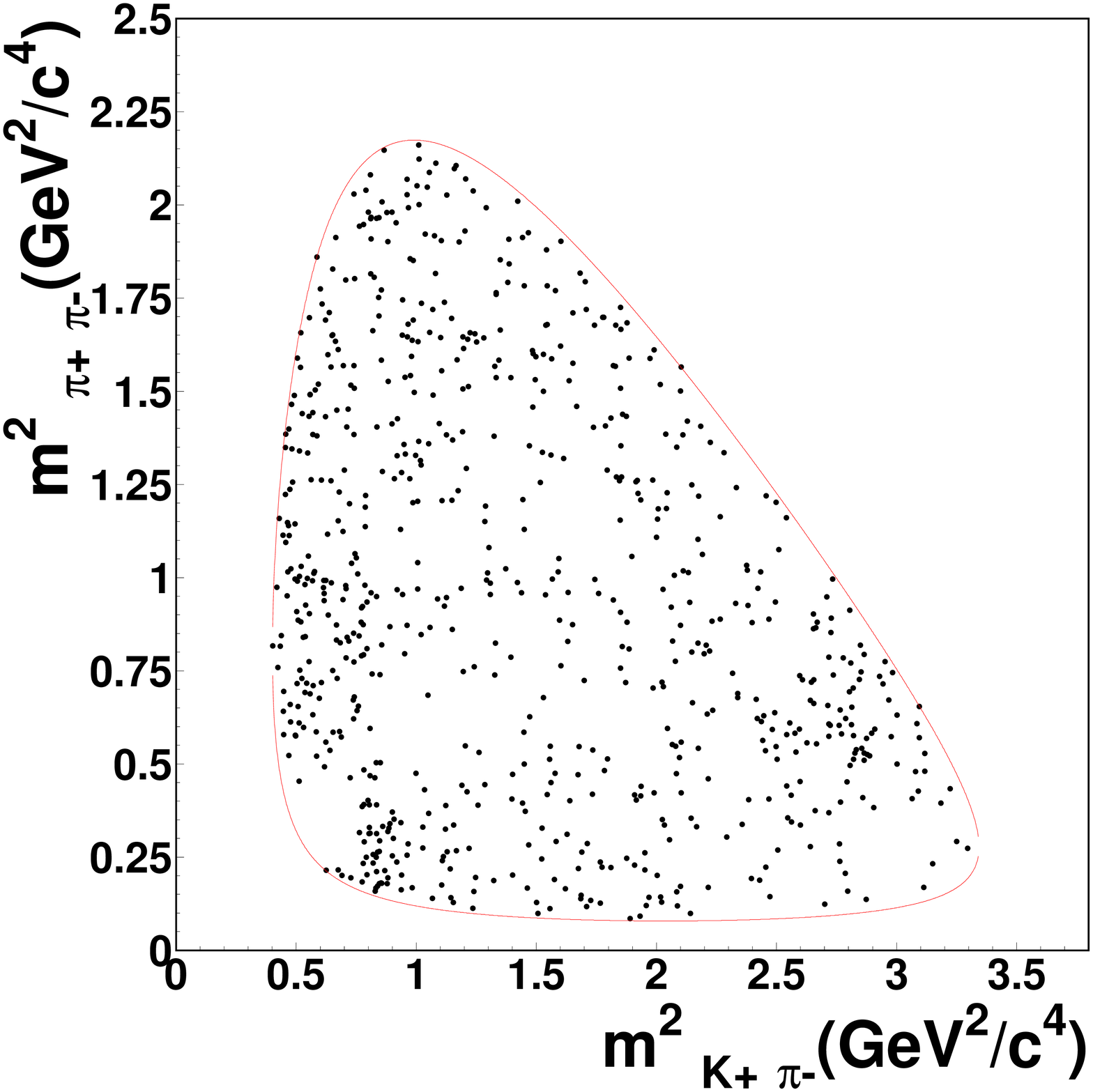}
  }
  \caption{ a) $D^+$ and b) $D^+_s$ Dalitz plots in $K^+ \pi^-$ and $\pi^+ \pi^-$ squared mass
  contributions.}
 \label{dalitz}
 \end{center}
\end{figure}

Events within $\pm 2 \sigma$ of the mass peak are used to perform the
Dalitz plot analysis for $D^+$ and $D^+_s \to K^+ \pi^+ \pi^-$; the $D^+$ and
$D^+_s$ Dalitz plots are shown in Fig.\ref{dalitz} a) and b), respectively.

\subsection{The decay amplitude}
For this analysis we use a formalization of the decay amplitude based on the
simple isobar model, which is viewed as an effective description of the
main dynamical features of the decay. More rigorous treatments, such as that
based on the \emph{K-matrix} formalism \cite{K_matrix}, are not viable for this
analysis because of the limited statistics of the samples, along with the large
number of free parameters necessary to account for the simultaneous presence of
both $\pi^+ \pi^-$ and $K^+ \pi^-$ resonances.

The decay matrix element for the analysis reported here is written as a coherent
sum of amplitudes
corresponding to a constant term for the uniform direct three-body decay and to
different resonant channels:
\begin{equation}
\mathcal{M} =  a_0 e^{i\delta_0}+ \sum_j a_j~ e^{i\delta_j}~B(abc|r)
\label{totamp}
\end{equation}

where $a,b$ and $c$ label the final-state particles.
$B(abc|r)=B(a,b|r)S(a,c)$ where $B(a,b|r)$ is the Breit-Wigner function
\begin{equation}
B(a,b|r) =\frac{F_DF_r}{M^2_r -M^2_{ab}-i\Gamma M_r}
\end{equation}

and $S(a,c)=1$ for a spin-0 resonance, $S(a,c) =-2{\bf c} \cdot {\bf a}$ for a
spin-1 resonance and $S(a,c) =2(|{\bf c}||{\bf a}|)^2(3\cos^2\theta^* -1)$ for
a spin-2 state. ${\bf a}$ and ${\bf c}$ are the three-momenta of particles
$a$ and $c$ measured in the $ab$ rest frame, and $\cos\theta^*={\bf c}\cdot {\bf a}/
|{\bf c}||{\bf a}|$.
The momentum-dependent form factors $F_D$ and $F_r$ represent the strong
coupling at each decay vertex. For each resonance of mass $M_r$ and spin $j$ we
use a width

\begin{equation}
\Gamma = \Gamma_r \left(\frac{p}{p_r}\right)^{2j+1}\frac{M_r}{M_{ab}}
\frac{F^2_r(p)}{F^2_r(p_r)}
\end{equation}

where $p$ is the decay three-momentum in the resonance rest frame and the
subscript $r$ denotes the on-shell values. The order of the particle labels is
important for defining the phase convention; here the first particle is the
opposite-sign one.

For the $f_0(980)$ a simple single-channel Breit-Wigner is used. The mass and
width values of $0.972\pm 0.002~\textrm{GeV}$ and $0.059 \pm
0.004~\textrm{GeV}$, 
respectively, are obtained
from a fit to our FOCUS Dalitz plot of the $D_s^+ \to \pi^+\pi^-\pi^+$ channel,
where the $f_0(980)$ is the dominant component. Given the questionable assumption
of a single channel Breit-Wigner, where a coupled-channel one should be more
properly used, we have checked that the final results do not change when a
Flatt\'e parametrization \cite{flatte} is used.
\subsection{The likelihood function and fitting procedure}

Following our previous analyses \cite{K_matrix,E687_ppp,E687_KKp} we perform a
maximum likelihood fit to the Dalitz plots to measure the coefficients of the
various decay amplitudes as well as their relative phases. The probability density
function is corrected for geometrical acceptance and reconstruction efficiency.
The shape of the background in the signal region is parametrized through an
incoherent sum of a polynomial function plus resonant Breit-Wigner components,
which are used to fit the Dalitz plot of mass sidebands. The number of
background events expected in the signal region is estimated through fits to the
$K^+\pi^+\pi^-$ mass spectrum. All background parameters are included as
additional fit parameters and tied to the results of the sideband fits through the
inclusion in the likelihood of a $\chi^2$ penalty term derived from the covariance
matrix of the sideband fit. The $D^+$ and $D_s^+$ samples are fitted with
likelihood functions $\mathcal{L}$ consisting of signal and background probability
densities. Checks for fitting procedure are made using Monte Carlo techniques and
all biases are found to be small compared to the statistical errors. The
systematic errors on our results are evaluated following the strategy already
explained for the branching ratio measurements, defining a \emph{split-sample} and 
a \emph{fit-variant} component. 
The assumption that the shape of the
background in the sideband is a good representation of the background in the
signal region potentially constitutes a source of
systematic error. We take into account this effect in the \emph{fit-variant} error, 
by varying the polynomial function degree
and adding/removing the Breit-Wigner terms, which are introduced to take into
account any feed-through from resonances in the background, and computing the
standard deviation of the different results.

In our Dalitz plot analysis we allow for the possibility of contributions from 
all known $K^+ \pi^-$ and
$\pi^+ \pi^-$ resonances\cite{pdg} and from a flat non-resonant contribution accounting
for the direct decay of the D mesons into three-body final states. The fit
parameters are amplitude coefficients $a_j$ and phases
$\delta_j$.\footnote{Coefficients and phases of the $\rho(770)$, which is a dominant
mode in both decays, are fixed to 1 and 0, respectively.} The general procedure,
adopted for all the fits reported here, consists of several successive steps in
order to eliminate contributions whose effects on our fit are marginal.
Contributions are removed if their amplitude coefficients, $a_0$ and $a_i$ of
Eq.~\ref{totamp} are less than $3\,\sigma$ significant \emph{and} the fit
confidence level increases due to the decreased number of degrees of freedom in
the fit. The fit confidence levels (C.L.) are evaluated with a
$\chi^2$ estimator over a Dalitz plot with bin size adaptively chosen to maintain
a minimum number of events in each bin. Once the minimal set of parameters is
determined, addition of each single contribution previously eliminated is reinstated
to verify if the C.L. improves; in this case the contribution is added in the
final set.

\subsection{Results}
\label{results}

The $D^+$ and $D^+_s$ Dalitz
plots present a different event intensity distribution. The $D^+$ appears highly
structured and dominated by $\rho(770)$ and $K^*(892)$.
The $D^+_s$ channel, instead, presents a much more complex structure diffused over
all the phase-space together with $\rho(770)$ and $K^*(892)$, whose main
characteristics are still visible. These general features are confirmed by our
fits.

\begin{figure}[!t]
 \begin{center}
  \includegraphics[width=1\textwidth]{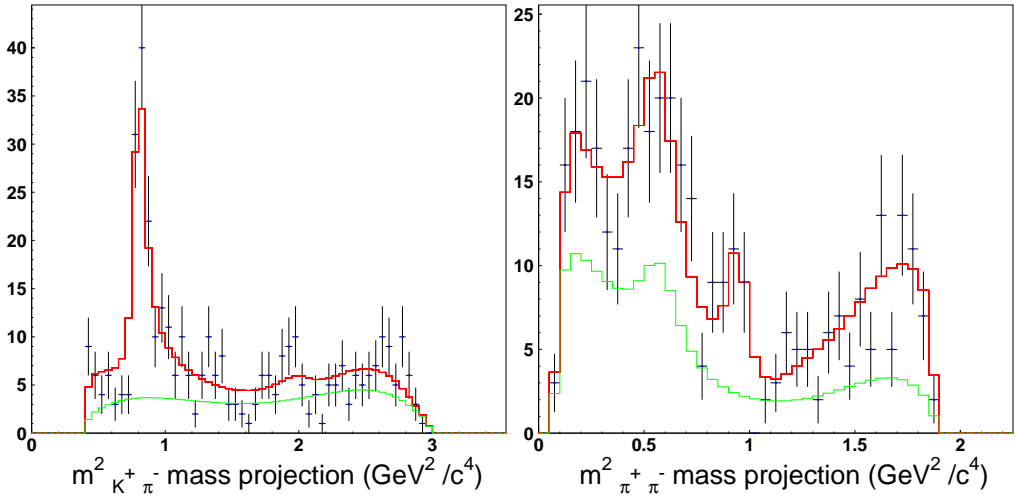}
  \caption{The Dalitz plot projections and, superimposed, the fit results for
  $D^+$. The background shape under the signal is also shown.}
  \label{fit_dp}
\end{center}
\end{figure}

In Table~\ref{tab_dalitz_dp}, fit fractions,\footnote{ The quoted fit fractions are
defined as the ratio between the intensity for a single amplitude integrated over
the Dalitz plot and that of the total amplitude with all the modes and
interferences present.} phases and coefficients of the various amplitudes
describing the $D^+ \to K^+\pi^+\pi^-$ decay are reported. 
\begin{table}[!t]
 \begin{center}
 \begin{tabular}{cccc}
 \hline
 \hline
 Decay channel & Fit fraction (\%) & Phase $\phi_j$ (degrees)  &
 Amplitude coefficient\\
 \hline
 \hline
 $\rho(770)K^+$   &  $39.43\pm7.87\pm8.15$    & $0$ (fixed)    &$1$ (fixed)\\
 \hline
 $K^*(892)\pi^+$  &  $52.20\pm6.84\pm6.38$	 & $-167.1\pm14.4\pm23.0$     &
 $1.151\pm0.173\pm0.161$\\
 \hline
 $f_0(980)K^+$    &  $8.92\pm3.33\pm4.12$	& $-134.5\pm31.4\pm41.9$
 &$0.476\pm0.111\pm0.143$\\
 \hline
 $K_2^*(1430)\pi^+$ &  $8.03\pm3.72\pm3.91$	& $54.4\pm38.3\pm20.9$
 &$0.451\pm0.125\pm0.129$\\
 \hline
 \hline
  C.L. = 9.2\%& $\chi^2 = 13.6$ & \multicolumn{2}{c} {d.o.f. = 21 (\#bins) - 13 (\#free
  parameters)}\\
 \hline 
 \hline
 \end{tabular}
 \caption{Dalitz plot fit results for the $D^+\to K^+ \pi^+ \pi^-$
 final state.}
 \label{tab_dalitz_dp}
 \end{center}
\end{table}
The vector resonances $\rho(770)$ and $K^*(892)$ account for about 90\% of the
$D^+$ decay fraction. Their relative phase difference, almost real, suggests a
marginal role for final state interactions in this decay. The C.L. of our
Dalitz plot fit is 9.2\%. The three-lobe helicity structure visible in the 
$D^+$ Dalitz plot justifies the presence of the tensor $K_2^*(1430)$, at more than three
sigma statistical significance, in the final set of resonances. The band of events
at about $(1 ~\textrm{GeV}/c^2)^2$ in the $ m^2_{\pi^+\pi^-}$ mass combination indicates the
presence of the scalar $f_0(980)$ in the decay at a four sigma significance level.
The absence in the $K^+\pi^-$ combination of scalar $K_0^*(1430)$, while the tensor
$K_2^*(1430)$ is present, seems to us a bit suspicious; further investigations
would required higher statistics. It is nevertheless interesting to observe
that the DCS decay $D^+ \to K^+\pi^+\pi^-$ is dominated by vector resonances with
no major role of rescattering effects.
The results of the fit on the two invariant mass squared projections
$m^2_{K\pi}$ and $m^2_{\pi\pi}$ are shown in Fig.~\ref{fit_dp}.

An amplitude analysis of this decay has been previously performed by the E791
experiment \cite{E791}, which described the decay with two resonant channels,
$\rho(770)K^+$ and $K^*(892)\pi^+$, plus a uniform non-resonant component, each
accounting for about $1/3$ of the decay fraction; the
non-resonant contribution seems to be better resolved in the present analysis in
additional resonant channels. 
We observe that the E791 $\rho(770)/K^*(892)$ relative phase difference\footnote{
This relative phase difference is free of phase
convention ambiguity.}
is about $ 0^{\circ}$ while our measurement is close 
to $180^{\circ}$. We also perform a fit of our sample with the same components as 
E791. The results still confirm 
the disagreement. 

In Table~\ref{risultati_ds_nok900}, the fit fractions, phases and coefficients
resulting from the $D_s^+$ fit are reported.
As for the $D^+$, the $\rho(770)$ and $K^*(892)$ vector resonances represent the
major contributions and cover about 60\% of the decay fraction; their phase shift
configuration is almost real for this decay as well. The description of the event
intensity all over the Dalitz plot requires three higher mass resonances, two in
$K^+\pi^-$ and one in $\pi^+\pi^-$, and a non-resonant term. More precisely the 
two $K^+\pi^-$ states are the scalar $K^*_0(1430)$ and the vector $K^*(1410)$, which 
are the lowest mass resonances besides the $K^*(892)$, and the $\pi^+\pi^-$ state 
is $\rho(1450)$, which is the second vector state in the $\rho$ series.
They account for an additional  30--40\%
resonant portion of the decay; a non-resonant contribution of about 15\% completes the
event description for this channel. The fit C.L. for the mixture of states
selected is 5.5\%.

\begin{figure}[!t]
 \begin{center}
  \includegraphics[width=1\textwidth]{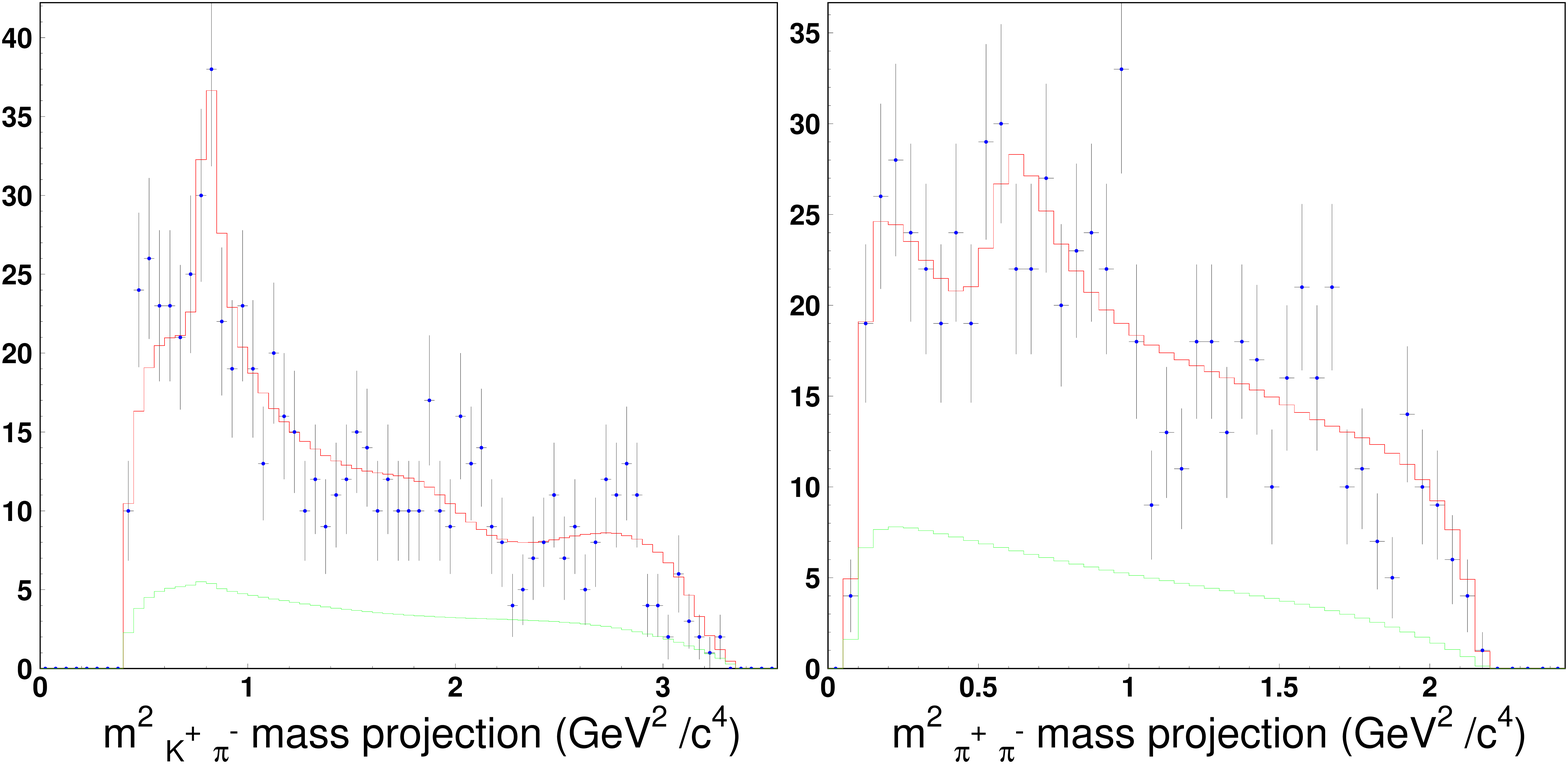}
  \caption{The Dalitz plot projections and, superimposed, the fit results for
  $D^+_s$. The background shape under the signal is also shown.}
  \label{fit_ds_sinr}
\end{center}
\end{figure}

This solution satisfactory reproduces the main features of the decay, as indicated
by the C.L., and shown in the two invariant mass squared projections $m^2_{K\pi}$
and $m^2_{\pi\pi}$ of the Dalitz plot in Fig.~\ref{fit_ds_sinr}. However
the absence of the $f_0(980)$ in the fit is a bit suspicious; an accumulation of
events at $(1 ~\textrm{GeV}/c^2)^2$ $~\pi^+\pi^-$ mass squared, to some extent visually 
recognizable in the Dalitz plot, would indeed suggest its selection in the resonance 
final set. 
On the other hand we know that the isobar model is too naive to describe more 
complex decays dynamics, which intervenes in the presence of the $K^+\pi^-$ and 
$\pi^+\pi^-$ S-waves states. More rigorous treatments, such as that based on the
\emph{K-matrix} model \cite{K_matrix}, will be necessary at higher statistics.
\begin{table}[!t]
 \begin{center}
 \begin{tabular}{cccc}
 \hline
 \hline
 Decay channel & Fit fraction (\%) & Phase $\phi_j$ (degrees) &
 Amplitude coefficient\\
 \hline
 \hline
 $\rho(770)K^+$   &  $38.83\pm5.31\pm2.61$    & $0$ (fixed)    &$1$ (fixed)\\
 \hline
 $K^*(892)\pi^+$  &  $21.64\pm3.21\pm1.14$     & $161.7\pm8.6\pm2.2$
 &$0.747\pm0.080\pm0.031$\\
 \hline
 $NR$    &  $15.88\pm4.92\pm1.53$      & $43.1\pm10.4\pm4.4$
 &$0.640\pm0.118\pm0.026$\\
 \hline
 $K^*(1410)\pi^+$   &  $18.82\pm4.03\pm1.22$    & $-34.8\pm12.1\pm4.3$
 &$0.696\pm0.097\pm0.025$\\
 \hline
 $K_0^*(1430)\pi^+$ &  $7.65\pm5.0\pm1.70$      & $59.3\pm19.5\pm13.2$
 &$0.444\pm0.141\pm0.060$\\
 \hline
 $\rho(1450)K^+$   &  $10.62\pm3.51\pm1.04$    & $-151.7\pm11.1\pm4.4$
 &$0.523\pm0.091\pm0.020$\\
 \hline
 \hline
  C.L. = 5.5\%& $\chi^2 = 38.5$ & \multicolumn{2}{c} {d.o.f. = 43 (\#bins) -  17 (\#free
  parameters)}\\
 \hline 
 \hline
 \end{tabular}
 \caption{Dalitz plot fit results for the $D^+_s\to K^+ \pi^+ \pi^-$
 final state.}
 \label{risultati_ds_nok900}
 \end{center}
\end{table}
The results of the fit on the squared invariant masses $m^2_{K\pi}$ and
$m^2_{\pi\pi}$ are shown in Fig.~\ref{fit_ds_sinr}.

The dominant systematic uncertainties on coefficients and phases for both $D^+$ and
$D^+_s$ decays come from the low/high momentum split and the fit variant component.


\section{Conclusions}

We have presented a study of the doubly and singly Cabibbo suppressed decays $D^+$
and $D^+_s \to K^+ \pi^+\pi^-$. Our measurements of $\Gamma(D^+\to
K^+\pi^+\pi^-)/\Gamma(D^+\to K^-\pi^+\pi^+)= 0.0065\pm 0.0008\pm
0.0004$ and $\Gamma(D^+_s\to K^+ \pi^+ \pi^-)/ \Gamma(D^+_s\to K^+
K^- \pi^+)=0.127 \pm 0.007\pm 0.014$ improve the statistical accuracy by
approximately a factor of 2 and 5 with respect to previous determinations.
 In particular the comparison of $(1/\tan^4\theta_\text{C})\times
\Gamma(D^+_{\text{DCS}})/\Gamma(D^+_{\text{CF}})=2.60\pm 0.32$ with the FOCUS lifetime ratio of
$\tau(D^+)/\tau(D^0)=2.538 \pm 0.023$ and the marginal role of FSI inferred by our
Dalitz plot analysis of this decay supports the interpretation that destructive
interference between spectator amplitudes with indistinguishable quarks in the CF
$D^+$ final state is responsible for the lifetime difference between $D^+$ and
$D^0$. The FOCUS collaboration has already analysed the doubly Cabibbo suppressed
decay of the neutral meson $D^0 \to K^+\pi^-$ \cite{focus_d0}; the study of the
DCS decay of the charged $D^+$ meson presented in this paper, free of any possible
uncertainty due to $D^0 \bar D^0$ mixing effects, provides complementary
information.
The amplitude analysis of $D^+$ and  $D^+_s \to K^+\pi^+\pi^-$ final
states have also been performed, providing the first amplitudes and phases for the
$D^+_s$ channel.
We have discussed the major achievements and indicated possible improvements for a better
decay dynamics interpretation at higher statistics.

We wish to acknowledge the assistance of the staffs of
Fermi National Accelerator Laboratory, the INFN of Italy, and the physics
departments of the collaborating institutions. This research was supported in
part by the US National Science Fundation, the US Department of Energy, the
Italian Istituto Nazionale di Fisica Nucleare and Ministero dell'Istruzione
dell'Universit\`a e della Ricerca, the Brazilian Conselho Nacional de
Desenvolvimento Cient{\'\i}fico e Tecnol\'ogico, CONACyT-M\'exico, the Korean
Ministry of Education, and the Korean Science and Engineering Foundation.

\end{document}